\documentclass[]{spie}  

\usepackage{amsmath,amsfonts,amssymb}
\usepackage{graphicx}
\usepackage[colorlinks=true, allcolors=blue]{hyperref}

\def\arcdeg{\mbox{$^\circ$}}%
\def\arcmin{\mbox{$^\prime$}}%
\newcommand{\fevv}{$^{\rm{55}}$Fe\ }

\title{The Advanced CCD Imaging Spectrometer on the Chandra X-ray Observatory: twenty-five years of on-orbit operation}

\author[a]{Catherine E.\ Grant}
\author[a]{Marshall W.\ Bautz}
\author[b]{Paul P.\ Plucinsky}
\author[a]{Peter G.\ Ford}
\affil[a]{Kavli Institute for Astrophysics and Space Research, Massachusetts Institute of Technology, Cambridge, MA, USA}
\affil[b]{Center for Astrophysics | Harvard \& Smithsonian, Cambridge, MA, USA}

\authorinfo{Further author information: \\
C.\ E.\ Grant: E-mail: cgrant@mit.edu}

\begin{document} 
\maketitle

\begin{abstract}
As the Advanced CCD Imaging Spectrometer (ACIS) on the Chandra X-ray Observatory completes a quarter century of on orbit operations, it continues to perform well and produce spectacular scientific results. The response of ACIS has evolved over the lifetime of the observatory due to radiation damage, molecular contamination, changing particle environment, and aging of the spacecraft in general. We present highlights from the instrument team's monitoring program and our expectations for the future of ACIS. Performance changes on ACIS continue to be manageable, and do not indicate any limitations on ACIS lifetime. We examine aspects of the design and operation of ACIS that have impacted its long lifetime with lessons learned for future instruments.
\end{abstract}


\section{INTRODUCTION}
\label{sec:intro}  
\begin{figure}
\begin{center}
\includegraphics[width=3in]{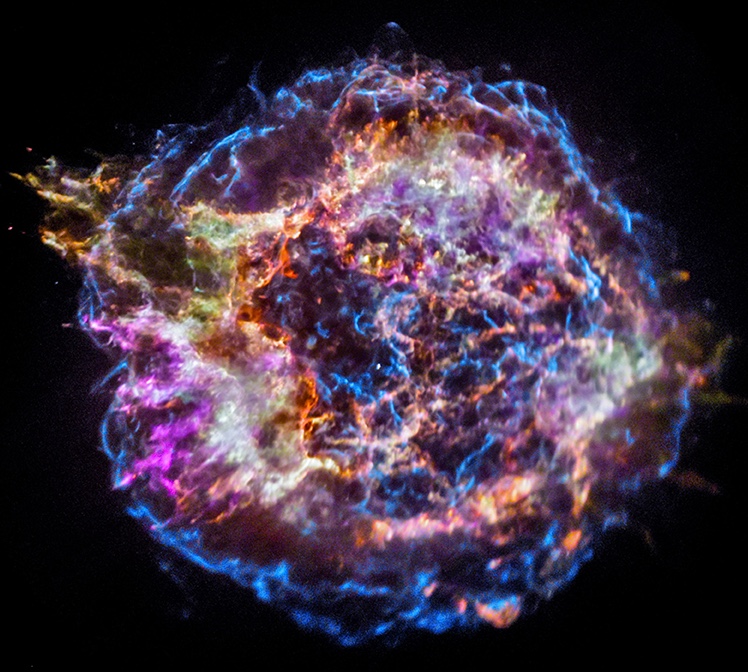}
\includegraphics[width=3.3in]{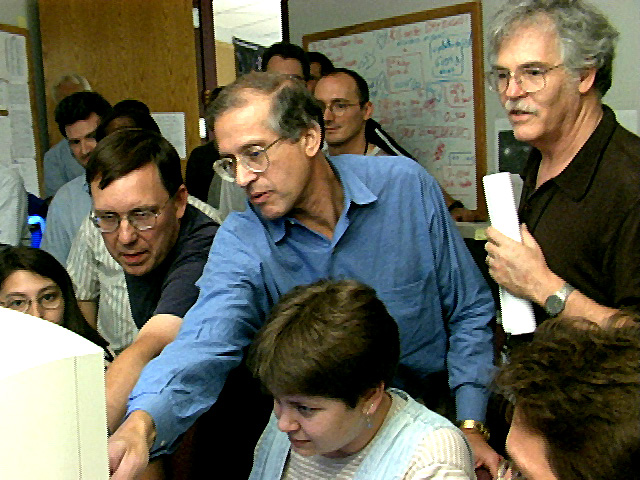}
\vspace*{0.1in}
\caption{(left) A Mega-second image of the young supernova remnant Cassiopeia A as imaged by Chandra and ACIS over a ten year period.  The image colors indicate the location of specific elements (Si, S, Ca, Fe) within the remnant determined by using the energy resolution of the CCD camera to resolve X-ray spectral features.  The small white point source in the center is the neutron star which had not been identified previous to Chandra. (right) The Chandra Operations Control Center as the first light observation of Cas A was underway. The central compact object was immediately apparent in the first images. (Image credit: NASA/CXC/SAO)}
\label{fig:casA}
\end{center}
\end{figure}

The Chandra X-ray Observatory, the third of NASA's great observatories in space, was launched shortly after midnight on July 23, 1999, aboard the space shuttle {\it Columbia}\cite{cha2}.  After a series of orbital maneuvers, Chandra reached its operational orbit, with initial 10,000-km perigee altitude, 140,000-km apogee altitude, and 28.5$^\circ$ inclination.  The first light observation of the supernova remnant Cassiopeia A immediately demonstrated the power of the new observatory by revealing in real-time the compact object at the core of the remnant (Figure~\ref{fig:casA}).

Chandra's High-Resolution Mirror Assembly (HRMA) provides exquisite sub-arcsecond X-ray imaging that was ground-breaking at launch and continues to be unmatched by any X-ray observatory in operation or in active development.  Two focal plane instruments, the Advanced CCD Imaging Spectrometer (ACIS) and the High-Resolution Camera (HRC) can be moved in and out of the telescope focus.  In addition, two grating assemblies can be moved in and out of the optical path: the High-Energy Transmission Gratings (HETG) and the Low-Energy Transmission Gratings (LETG). ACIS provides moderate spectral resolution imaging and serves as the readout detector for the HETG, while HRC has slightly better spatial resolution, minimal spectral resolution, and is the readout detector for the LETG. For an overview of the mission and more details about the instrument complement, see the Chandra Proposers' Observatory Guide.\footnote{https://cxc.cfa.harvard.edu/proposer/POG/} ACIS is the workhorse for Chandra science with more than 90\% of the observation time.

With the exception of rare spacecraft anomalies and radiation shutdowns, Chandra has been in continuous operation for nearly twenty-five years.  Chandra's orbit allows for high observing efficiency of approximately 70\%, with the remaining time spent transiting Earth's radiation belts.  Twenty-five years later, the observing efficiency remains close to the theoretical maximum.  Demand for Chandra observations is high, with steady oversubscription factors of 4-6.  Refereed publications are expected to exceed 10,000 during this 25th anniversary year.  On average, PIs receive their data within two days of the end of an observation and have access to a suite of specialized software tools and threads to help analyze their data.

All of this has resulted in spectacular scientific results too numerous to be listed here. Chandra is particularly well suited to studying crowded fields, like the Galactic Center and star forming regions, and complicated spatial structures such as sloshing and feedback in galaxy clusters. A growing number of studies make use of Chandra's long lifetime and high spatial resolution to map tiny motions of supernova remnants and other nearby objects.  

The response of ACIS has evolved over the lifetime of the observatory due to radiation damage, contamination accumulation, and degradation of thermal control. This has resulted in some loss of low-energy effective area and a general need for careful calibration in the temporal, spatial, and spectral domains, and as a function of the focal plane temperature. The ACIS instrument team maintains an ongoing monitoring program, has developed software and calibration products to improve instrument response, and flight software patches that further protect the instrument to help ensure a long lifetime. 

In this paper, we discuss some aspects of the twenty-five years of Chandra on-orbit operation from the perspective of the ACIS instrument.  We begin with a more detailed description of the instrument in Section~\ref{sect:acis}, then a brief summary of current status and a deeper dive into selected topics in Section~\ref{sect:status}. Section~\ref{sect:lessons} outlines a few lessons learned before a summary in Section~\ref{sect:sum}.

\section{Detailed Description of the Instrument}
\label{sect:acis}

\begin{figure}
\begin{center}
\includegraphics[width=5.9in]{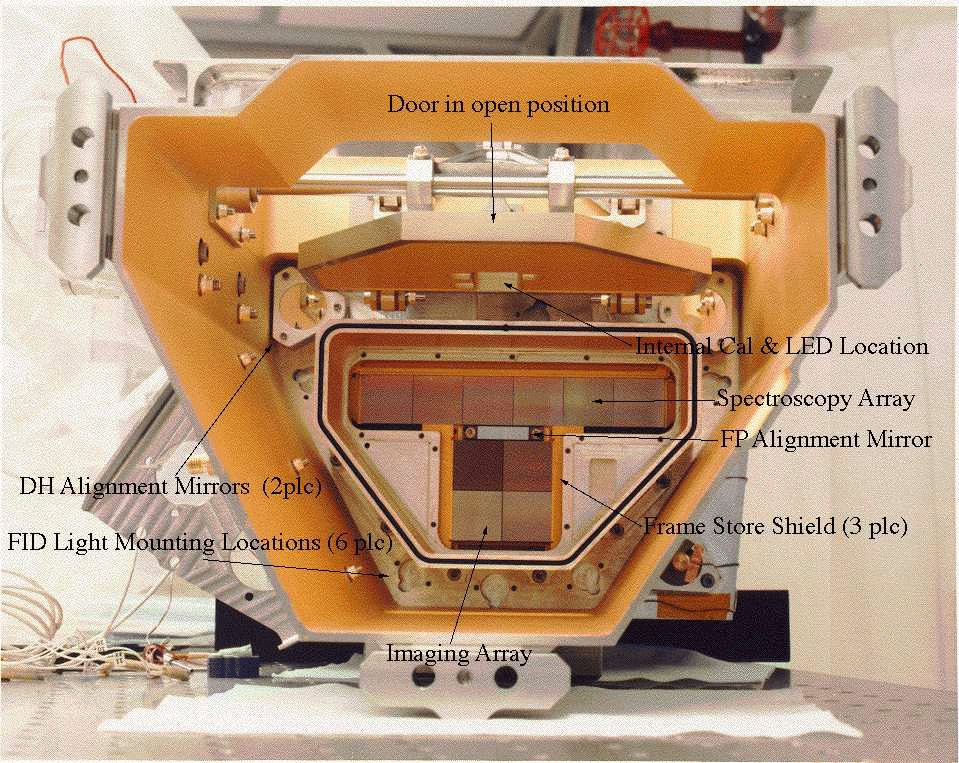}
\includegraphics[width=3.5in]{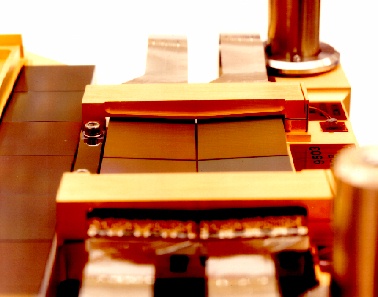}
\vspace*{0.1in}
\caption{(top) A photograph of the ACIS focal plane as it would appear looking down the optic axis. The $2\times2$ CCD square Imaging Array and six CCD linear Spectroscopy Array are visible, as are various structures that provide support and thermal functions. In flight, the CCDs are covered by an Optical Blocking Filter. (bottom) A side view of the focal plane shows the gold coated framestore covers and how the CCDs are tilted to best match the focal surface.}
\label{fig:ACISpic}
\end{center}
\end{figure}

The ACIS focal plane consists of two arrays.  The first, ACIS-I, is 2 $\times$ 2 CCDs tilted to best match the focal surface of the HRMA for highest spatial resolution.  The second, ACIS-S, intended as a readout for the HETG, is a linear 6 CCD array tilted to match the Rowland circle of the gratings.  Eight of the CCDs are front-illuminated with two back-illuminated CCDs in ACIS-S: one at the ACIS-S aimpoint and one further along the array where lower energy X-rays are dispersed.   In practice, ACIS-S has also been used for imaging smaller fields, as the response of the BI CCD is desirably for some investigations. Photos of the ACIS focal plane are shown in Figure~\ref{fig:ACISpic} and more characteristics are listed in Table~\ref{tab:ACISchar}.  ACIS was built by MIT and Lockheed Martin under the direction of PI Gordon Garmire from Penn State University (now at the Huntingdon Institute for X-ray Astronomy).

\begin{table}[t]
\caption{ACIS characteristics \label{tab:ACISchar}}
\begin{center}       
\begin{tabular}{|l|l|}
\hline\hline
Detectors & 10 frame-transfer X-ray CCDs \\ 
&\hspace{1cm}8 front-illuminated\\
&\hspace{1cm}2 back-illuminated\\\hline
CCD format &1024 $\times$ 1024 pixel image array \\ \hline
Image area pixel size &24 $\times$ 24 $\mu$m \\\hline
Depletion depth &$\sim$70 $\mu$m (FI), $\sim$45 $\mu$m (BI fully depleted) \\\hline
Output ports &on-chip MOSFET, 4 / CCD \\ \hline

Focal plane arrays: & \\ 
\hspace{1cm}ACIS-I &2 $\times$ 2 CCD array \\ 
 &tangent to mirror focal surface\\ 
 &16.9\arcmin $\times$ 16.9\arcmin\\\hline
\hspace{1cm}ACIS-S &1 $\times$ 6 CCD array \\
&tangent to grating Rowland circle\\ 
&8.4\arcmin $\times$ 51.1\arcmin \\\hline

Focal plane temperature    & $-120$\arcdeg C setpoint \\
&$-120$\arcdeg C to $-105$\arcdeg C current allowed range \\ \hline
Optical blocking filter (Al/Polyimide/Al) &120 nm / 200 nm / 40 nm (ACIS-I)\\ 
&100 nm / 200 nm / 30 nm (ACIS-S)\\\hline

System noise & 2--3 e- RMS\\\hline
Nominal frame time & 3.2 sec (full frame)\\ \hline
Allowable frame times & 0.1 to 10.0 sec \\ \hline
Maximum serial rate &100 kHz (10 $\mu$sec / pixel)\\\hline
Image-to-framestore parallel rate &25 kHz (40 $\mu$sec / row) \\\hline
Instrument output data rate &24 kbps \\\hline
\hline
\end{tabular}
\end{center}
\end{table}

\begin{figure}
\begin{center}
\includegraphics[width=5in]{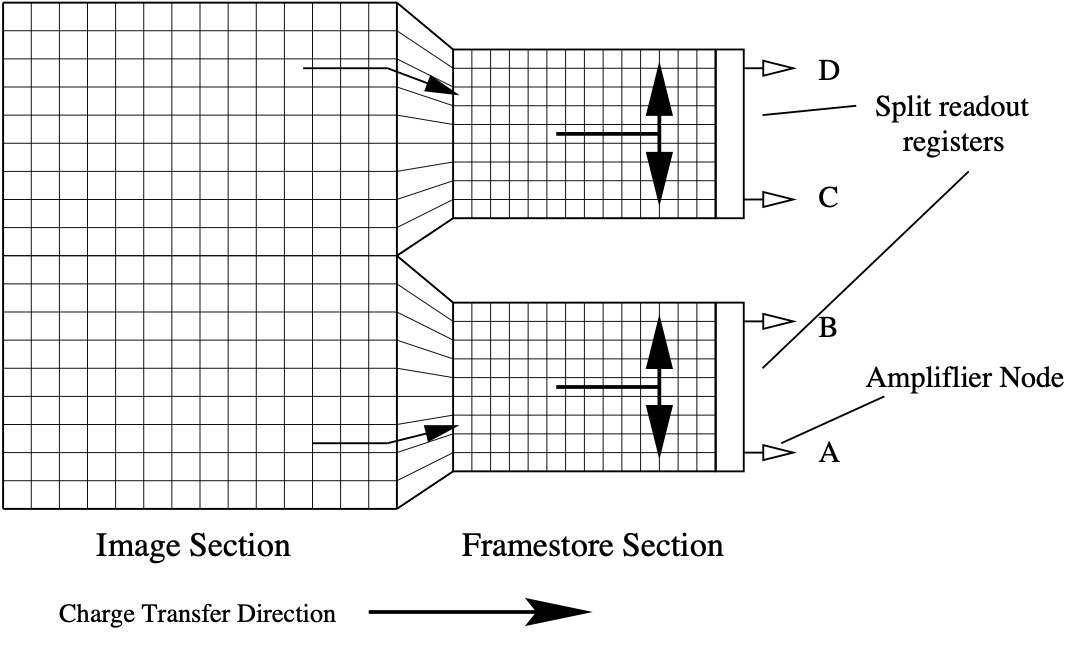}
\caption{A schematic drawing of an ACIS CCD showing the transfer direction through the image and framestore sections of the CCD to the serial register.  Charge from each half of the CCD can then be clocked out entirely to the node at one end or the other or can be split in half to use all four amplifier nodes.}
\label{fig:ACISdiag}
\end{center}
\end{figure}

The ACIS CCDs are CCID-17 frame transfer devices designed and manufactured for Chandra  by MIT Lincoln Lab, out of high-resistivity float-zone Silicon.  Each CCD has 1024 $\times$ 1024 24-$\mu$m pixels in the imaging array and three phase charge transfer through triple layer polysilicon transfer gates.  The FI CCDs have a depletion depth of $\sim$70~$\mu$m and the BI CCDs are fully depleted with a thickness of $\sim$45~$\mu$m. The buried channel includes an additional 2-$\mu$m trough implant to mitigate radiation damage. Figure~\ref{fig:ACISdiag} is a schematic diagram of an ACIS CCD showing the transfer direction and readout path.  Each CCD has four single-stage on-chip MOSFET readout amplifiers.  Fastest frame time is achieved by utilizing all four, but it is also possible to clock each half of the framestore out through one end or the other, if desired.  The framestore is covered by 3~mm of gold coated aluminum.

ACIS runs in photon-counting mode, with frametimes of $\sim$3~seconds. On-board processing identifies potential X-ray event candidates and telemeters the subset most likely to be X-rays and not particles to the ground. Event candidates are identified as pixels that are local maxima and have a pulseheight above an event threshold.  On-board filtering removes events with pulseheights above an upper threshold and event morphologies that are more likely to be charged particle events. While the removed events themselves are not telemetered, the instrument does keep an accounting of the numbers of events removed and why. We use one of these counters, the high energy reject rate, as a proxy for the particle rate seen by the spacecraft (Section~\ref{sect:bkg}) and another of the counters, number of pixels above the event threshold, as part of the on-board radiation monitor (Section~\ref{sect:txings}) .

ACIS offers many selectable observing parameters, including the choice of CCDs, restrictive energy filters, and reporting additional information from neighboring pixels in the telemetry stream.  Most options include trade-offs between faster readout, spatial coverage, and telemetry limitations. The choices and trade-offs are best described in the Chandra Proposers Observatory Guide.\footnote{https://cxc.cfa.harvard.edu/proposer/POG/}  To date, 1385 unique ACIS configurations have been built in response to observer requests and are available for use.

The focal plane CCDs and the camera body are passively cooled via two large radiators protected from direct sunlight, then maintained at the desired temperature with heaters.  Since January 2000, the focal plane temperature setpoint has been $-120\arcdeg$C.  As the spacecraft has aged and the multi-layer insulation (MLI) has degraded, temperatures of all spacecraft subsystems have increased.  The implications for ACIS will be discussed further in Section~\ref{sect:temp}.

More details of the ACIS CCDs and the ACIS focal plane can be found in Ref.~\citenum{ACISCCD} and \citenum{GarmireACIS}, respectively.

\subsection{External Calibration Source}
\label{sect:ecs}

A key element of ACIS on-orbit operations and calibration has been the External Calibration Source (ECS), which fully illuminates all ten CCDs while ACIS is in the stowed position. Since the discovery of the initial radiation damage (further discussed in Section~\ref{sect:cti}), a continuing series of observations of the ECS have been undertaken--just before and after the instruments are safed for perigee passages--to monitor the performance of the ACIS CCDs.  

\begin{figure}
\begin{center}
\includegraphics[height=3.5in]{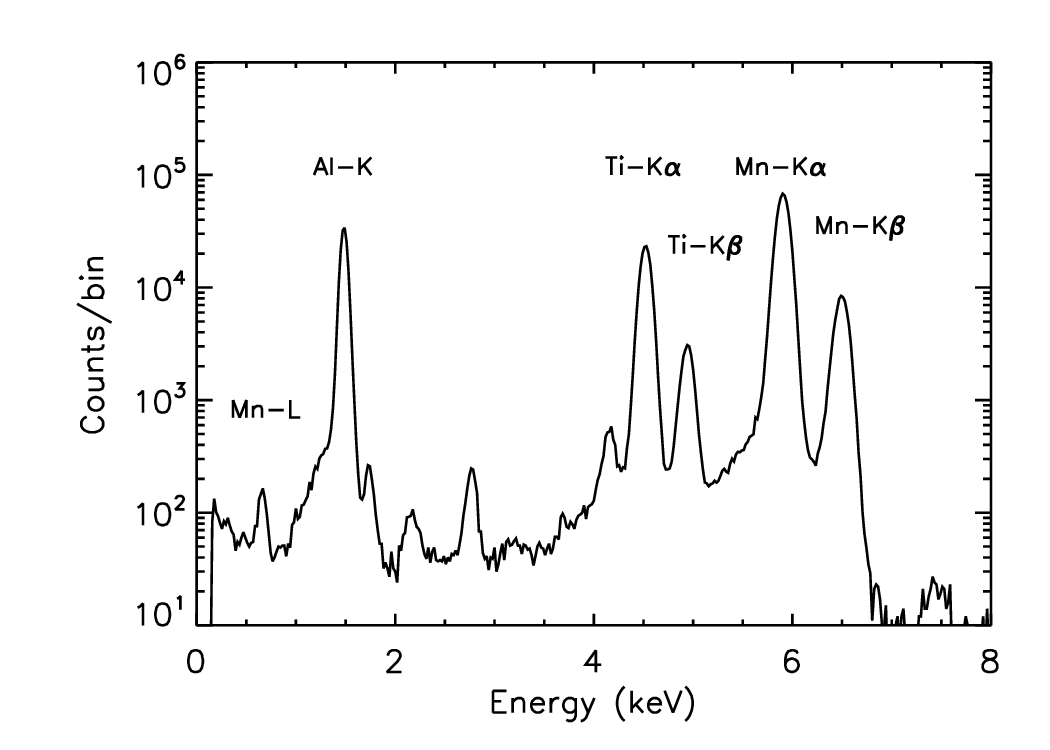}
\caption{An on-orbit spectrum of the ACIS External Calibration Source on an ACIS-I FI CCD.  Only data from the first 256 rows are included to minimize CTI degradation.  Lines used for calibration and monitoring purposes are labelled.  The remaining features are instrumental.}
\label{fig:spec}
\end{center}
\end{figure}

The ECS consists of three radioactive \fevv sources, two of which include fluoresence targets.  The ECS produces strong lines of Mn-K$\alpha$ and Mn-K$\beta$ at 5.9~keV and 6.5~keV and a much weaker complex of unresolved Mn-L lines at $\sim$700 eV.  Fluorescence targets add lines from Al-K at 1.5~keV and Ti-K$\alpha$ and Ti-K$\beta$ at 4.5~keV and 4.9~keV.  The access to multiple energies has proven invaluable in studying gain changes on short time-scales and the energy and spatial dependence of charge transfer inefficiency, as well as the change in low-energy efficiency due to molecular contamination.  Figure~\ref{fig:spec} shows the spectrum of the ACIS External Calibration Source and labels the important features.

The frequent cadence of ECS observations, roughly twice every three days, has allowed for close monitoring of ongoing radiation damage and rapid investigation of potential instrument anomalies, while still providing a long baseline for detailed studies of performance changes. In addition to the regular ECS observations near perigee, much longer exposures are often undertaken during observatory shutdowns due to solar storms and other spacecraft anomalies.

\begin{figure}
\begin{center}
\includegraphics[height=3.5in]{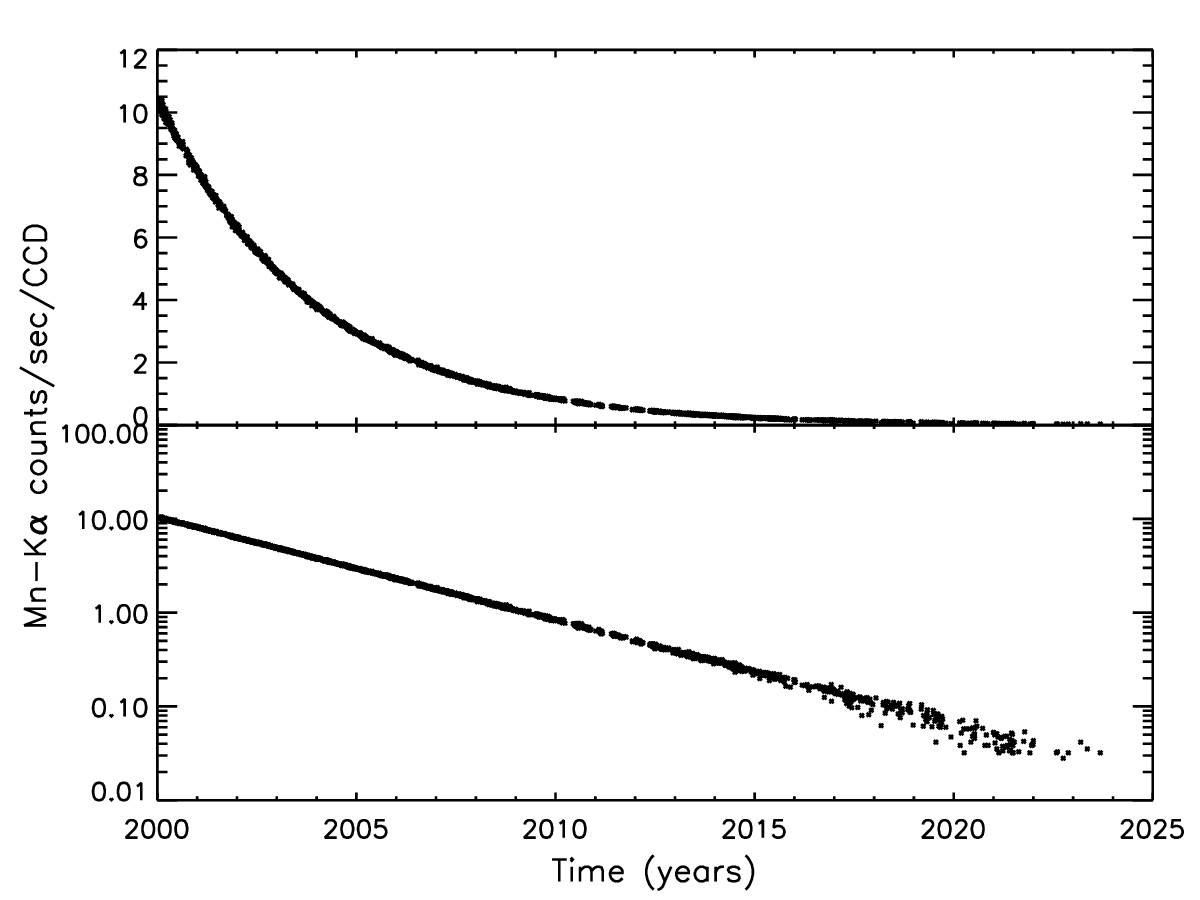}
\caption{The decline in the count rate of the ACIS ECS due to radioactive decay of the \fevv source. The reduction in the number of data points in recent times is due to a combination of fewer cold ECS data sets and poor fits due to low counts.}
\label{fig:kdecay}
\end{center}
\end{figure}

\fevv has a half-life of $\sim$2.7~years. Figure~\ref{fig:kdecay} shows the decline in the count rate of the Mn-K$\alpha$ line throughout the Chandra mission due to the decay of the calibration source. Clearly over the twenty-five year lifetime of Chandra, the count rate from the ECS has dropped substantially, requiring more thoughtful analysis and some reduction in the ability to simultaneously measure spatial and temporal effects.  The weakest line from the low energy Mn-L complex at $\sim$700~eV was essentially lost in the underlying background spectrum after 2010.  The stronger lines remain well measured and should continue to provide valuable calibration and monitoring data into the future with the appropriate spatial and temporal binning.  

The long Chandra lifetime allows us to measure the \fevv half-life directly from the ACIS ECS data.  A fit to this data yields a half-life of $2.740 \pm 0.005$~years, which agrees well with the often quoted value in the literature of $2.737 \pm 0.011$~years\cite{Fe55decay}.

\section{Instrument Status and On-orbit Performance Evolution}
\label{sect:status}

ACIS continues to perform superbly with no limitations on its future operation for many years.  All ten CCDs and forty readout chains are fully functional with minimal increase in cosmetic defects due to radiation damage.  The system readout noise from all chains remains close to that at launch.  All the instrument electronics and power supplies are nominal, operating on the original primary units with no failures.  Redundant electronics remain available, should they be needed.  The CCD output chains are multiplexed to the six front-end processing chains, so the loss of one front-end board would only mean a reduction in the number of available CCDs rather than the loss of a specific CCD.

There is no evidence of light leaks through the Optical Blocking Filter (OBF) and no indication of damage due to micrometeoroid impact on the filter or the CCDs.  There has been a continuing gradual deposition of molecular contamination on the OBF which is further discussed in Section~\ref{sect:contam}.

The ACIS flight software has been patched eleven times since launch including bug fixes, improvements to on-board processing, and enhancements to the flight software capabilities. Some of these patches will be discussed in Section~\ref{sect:txings}. The ACIS boot code stored in an on-board EEPROM is checked monthly for data corruption and has remained fault-free.

The only failures in the ACIS instrument to date are to two temperature sensors, one on the power supply box (in 2016) and one on the cold radiator (in 2023) which intermittently read open circuits. Neither is required for on-going thermal modeling and  other thermal sensors can be used for health and safety monitoring. The team has concentrated some effort on identifying temperature sensors that \textbf{are} necessary for operations and how their function might be duplicated, were they to fail in the future. 

\subsection{Charge Transfer Inefficiency}
\label{sect:cti}

Much of the early history of ACIS performance concerns charge transfer inefficiency (CTI).
Chandra's highly elliptical 2.7-day orbit exposes it to a wide range of particle environments, and transits the Earth's radiation belts at perigee.  

Soon after launch it was discovered that the FI CCDs had suffered radiation damage from exposure to soft protons scattered off the Observatory’s grazing-incidence optics during passages through the Earth’s radiation belts.\cite{PrigozhinCTI} The BI CCDs were unaffected. Since mid-September 1999, ACIS has been protected from the protons by moving the CCDs to the stowed position during radiation belt passages and there has been an ongoing effort to prevent further damage and to develop hardware and software strategies to mitigate the effects of charge transfer inefficiency on data analysis.\cite{TownsleyCTIcorr,GrantCTIcorr}

Our primary measure of radiation damage on the CCDs is charge transfer inefficiency (CTI), the fraction of charge lost during pixel-to-pixel transfer. The eight front- illuminated CCDs had essentially no CTI before launch (CTI $< 10^{-6}$), but are strongly sensitive to radiation damage from low energy protons ($\sim$100~keV) which preferentially create traps in the buried transfer channel. After eight radiation belt transits the FI CCD CTI had increased to $\sim 10^{-4}$. The framestore covers, 2.5-mm of gold coated aluminum, are thick enough to stop this radiation, so the initial damage was limited to the imaging area of the FI CCDs. Radiation damage from low-energy protons is now minimized by moving the ACIS detector away from the aimpoint of the observatory during passages through the Earth’s particle belts and during solar storms. Continuing exposure to both low and high energy particles over the lifetime of the mission slowly degrades the CTI further.\cite{OdellCTI,GrantCTI} The two back-illuminated CCDs (ACIS-S1,S3) had measurable CTI before launch in both the imaging and framestore areas (CTI $\sim 10^{-5}$) and in the serial transfer array. Owing to their much deeper charge-transfer channel, BI CCDs are insensitive to damage by the low-energy protons that damage FI devices.

\begin{figure}
\begin{center}
\includegraphics[height=3.5in]{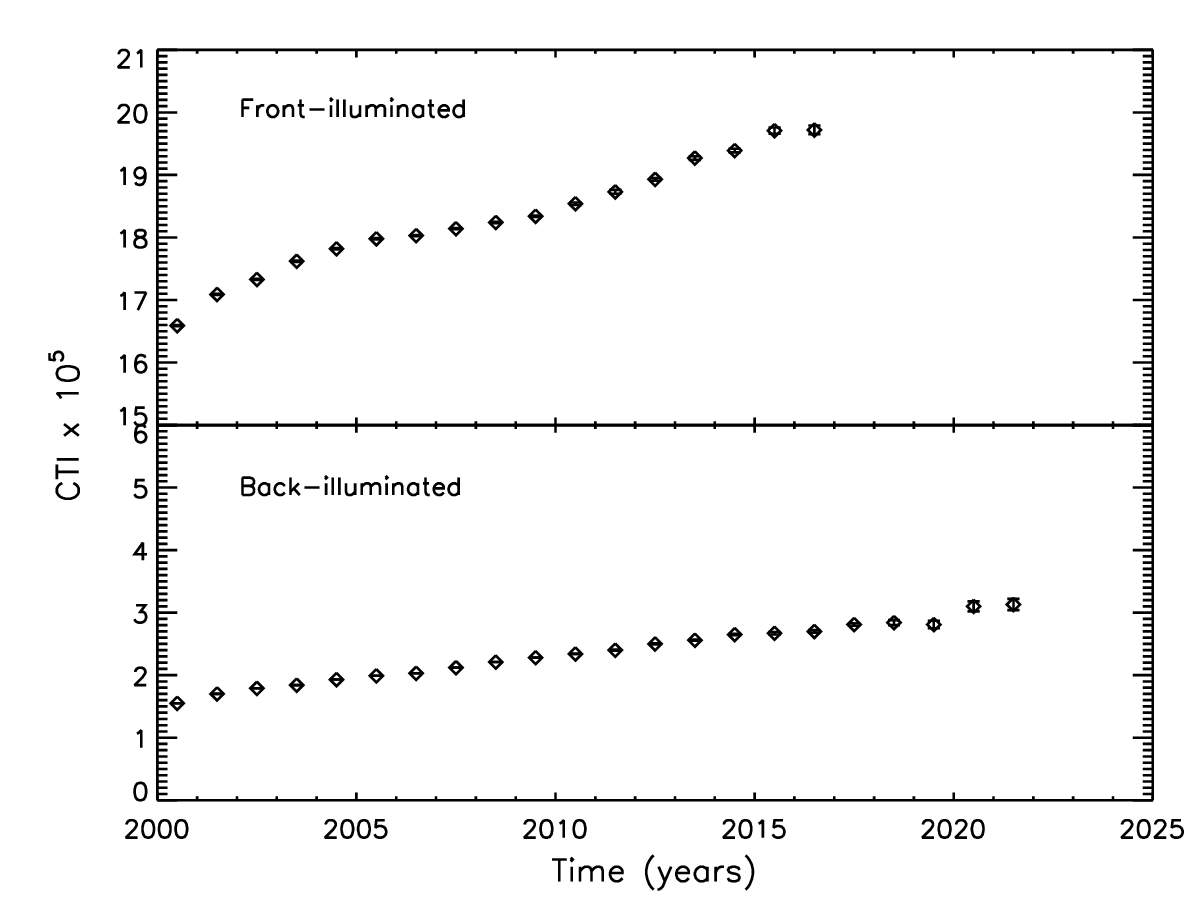}
\caption{Parallel CTI measured at 5.9~keV and -120\arcdeg C for a front-illuminated and back-illuminated ACIS CCD. The FI  CCDs have higher CTI, due to the radiation belt transits at the start of the mission.  Both types of CCDs show a gradual increase in CTI with time. }
\label{fig:cti}
\end{center}
\end{figure}

Figure~\ref{fig:cti} shows the parallel CTI from January 2000 onward, measured using the 5.9~keV line of the ECS after applying corrections for temperature dependence and sacrificial charge due to the particle background, and grouping the data into year-long bins. The lack of data in the last few years is due to a combination of factors: the weakening \fevv source, operational changes that reduce the time dedicated to these measurements, and an increase in high focal plane temperatures.

CTI increases smoothly and roughly linearly with time. The FI CCDs have much higher CTI than the BI CCDs, due to the initial radiation damage from low energy protons. The CTI increase of the FI CCDs ($\delta \mathrm{CTI} \sim 2 \times 10^{-6}$/yr) is higher than that of the BI CCD ($\delta \mathrm{CTI} \sim 1 \times 10^{-6}$/yr) which indicates that the accumulated radiation damage must include a mixture of particle energies, including the low energy protons which preferentially damage the FI CCDs but not the BI CCDs.

\subsection{Read Noise}

\begin{figure}
\begin{center}
\includegraphics[height=3.5in]{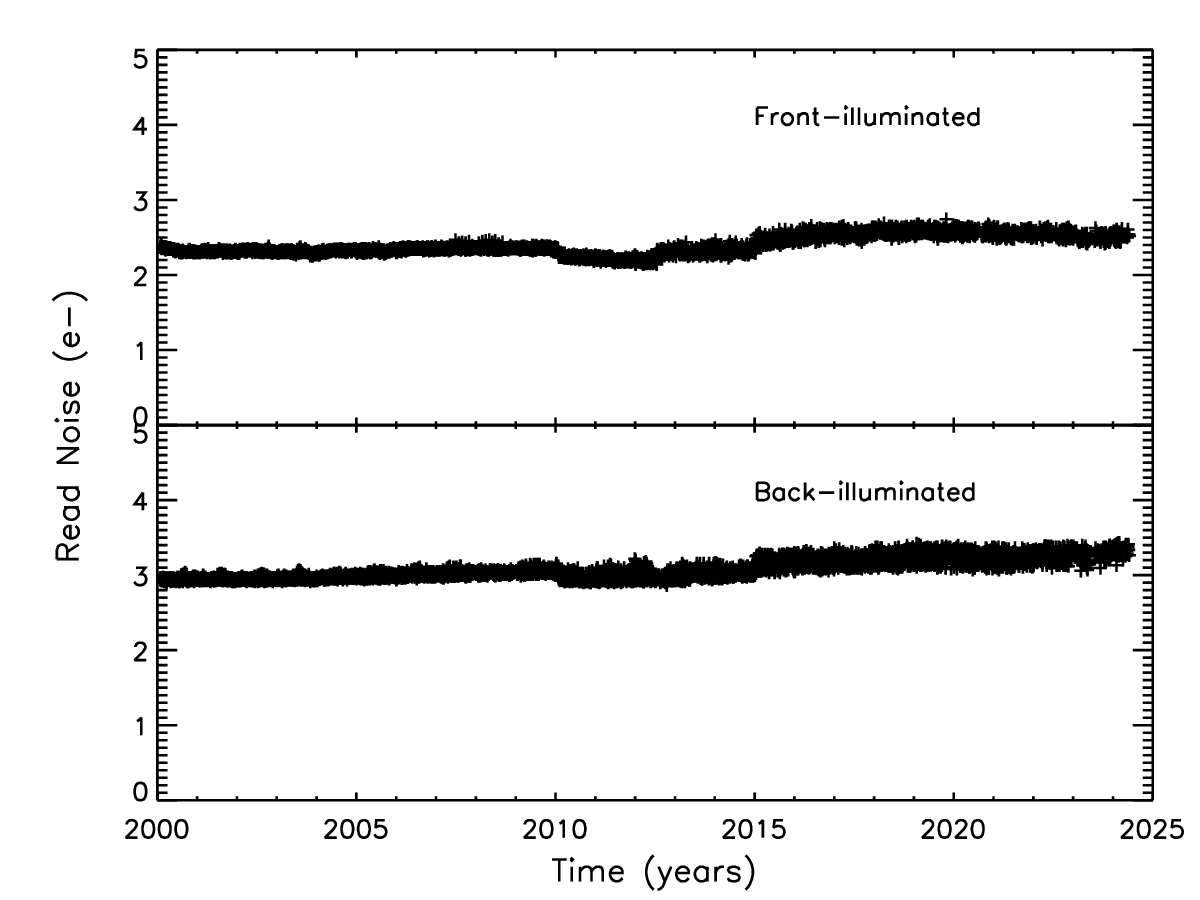}
\caption{The total system read noise of the front-illuminated I-array aimpoint CCD and the back-illuminated S-array aimpoint CCD as a function of time. There has been very little change in the total system noise since launch. Very small step changes are due to known instrument anomalies and operational changes.}
\label{fig:noise}
\end{center}
\end{figure}

Before launch, the ACIS CCDs had read-out noise less than 2 e- RMS, with a total system noise for the entire signal chain of 2-3 e- RMS.  Since launch there has been very little change in the system noise even after twenty-five years of exposure to a space environment.  Figure~\ref{fig:noise} shows the total system read noise on the front-illuminated CCD at the I-array aimpoint and the back-illuminated CCD at the S-array aimpoint since January 2000.  This was measured by fitting a Gaussian to the corner pixels of the 3$\times$3 event islands.  Since the corner pixels go through the same process on-board as the X-ray events, the measured width of the corner pixel Gaussian is representative of the noise seem by X-ray events including contributions from the entire readout and processing chain. The back-illuminated device was known to have higher noise before launch.  Both devices show a negligible increase of $\sim$ 0.3 e- over twenty-five years that is not a significant degradation of instrument performance.

\subsection{Temperature Control}
\label{sect:temp}

The ACIS focal plane is passively cooled via a radiator and then maintained at a set temperature via a heater.  In January 2000, that set point was $-120\arcdeg$C.  In subsequent years, the multi-layer insulation (MLI) of the spacecraft slowly degraded, and temperatures throughout the spacecraft began to rise.  ACIS focal plane temperatures are sensitive to the spacecraft pitch and roll, and sometimes receive extra heating around perigee due to earthshine on the radiator.

\begin{figure}[t!]
\begin{center}
\includegraphics[width=6.5in]{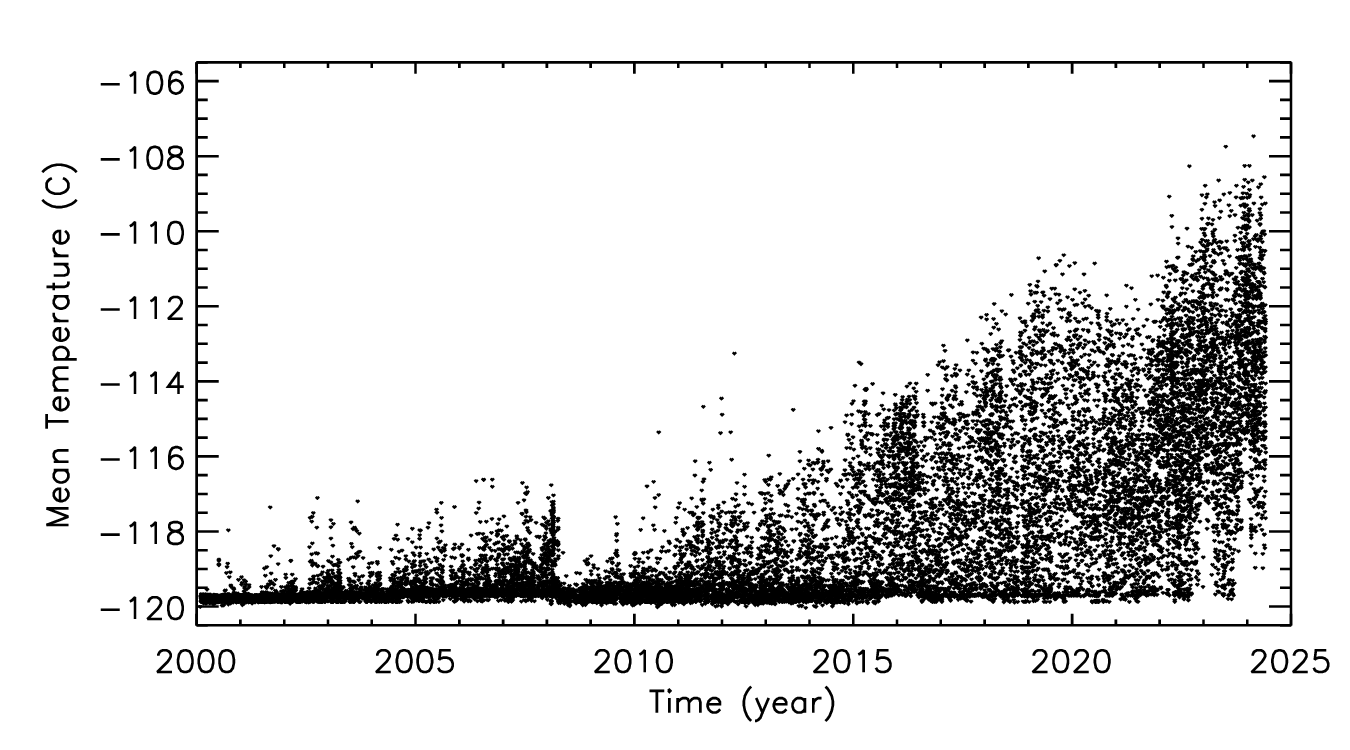}
\caption{Time dependence of focal plane temperature since January 2000.  Each data point is a single science observation. The nominal set point of the ACIS focal plane is $-120\arcdeg$C, but increasingly drifts to higher temperatures.}
\label{fig:temp}
\end{center}
\end{figure}

Figure~\ref{fig:temp} shows the time dependence of focal plane temperature. Each data point is the mean temperature of a single observing run.  It is clear that the frequency and magnitude of high focal plane temperatures have both increased with time.   In the past few years very little time is spent at the nominal setpoint value.  The radiator itself does not appear to have degraded, as tests have shown that given the proper spacecraft pitch angle and enough time, the temperature can drop well below $-120\arcdeg$C.  The issue is rather that the thermal limits of other spacecraft subsystems require pitch angles that are less favorable for cold ACIS focal plane temperatures.

The radiation induced CTI on ACIS, especially for the FI CCDs, is temperature dependent, which means that gain and spectral resolution both degrade with increasing temperature.\cite{CTItemp} Best performance for the FI CCDs requires a cold focal plane.  The calibration team has set limits on the maximum temperatures for good calibration.  In addition, the ACIS CTI correction software has a temperature component to help mitigate the dependence.  

\subsection{Contamination}
\label{sect:contam}

\begin{figure}[t!]
\begin{center}
\includegraphics[width=4in]{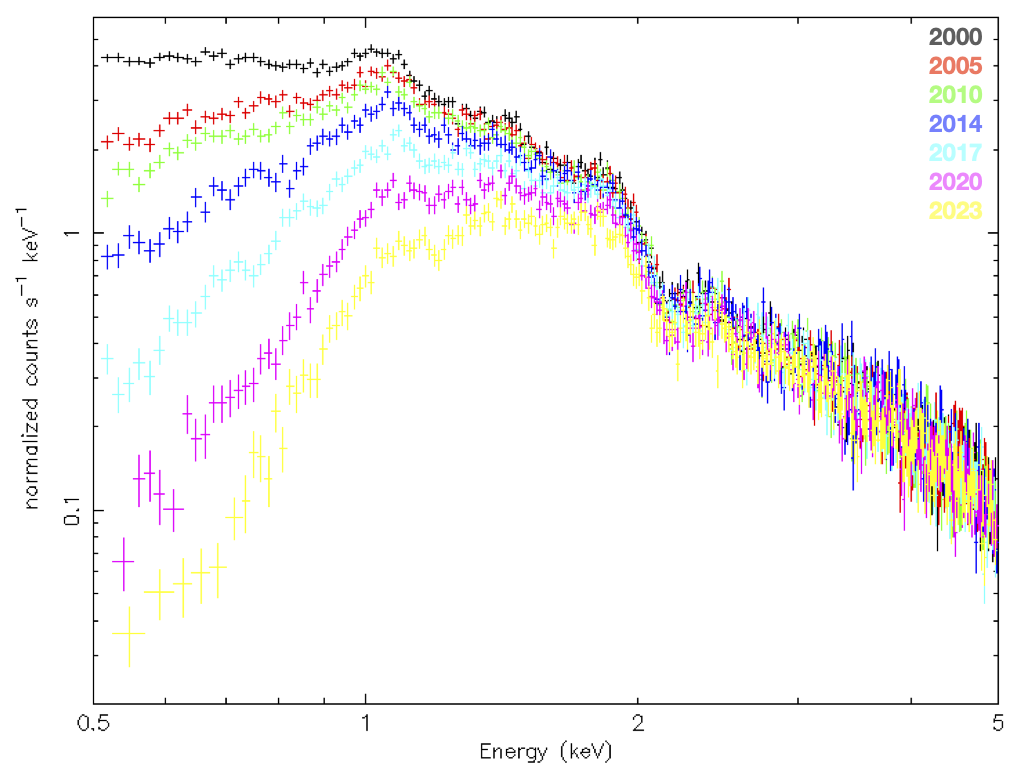}
\caption{Observations of a galaxy cluster from 2000 to 2023 showing the decrease in low energy effective area due to contamination accumulation.(Source: Chandra Observatory Proposers Guide)}
\label{fig:contam}
\end{center}
\end{figure}

Within a few years of launch it was clear that there was excess absorption from molecular contamination somewhere in the system.  The spatial distribution of the contaminant matched the temperature profile of the OBF, thicker/colder at the edges, thinner/warmer in the center, so the contaminant seems likely to be on the OBF, rather than on the CCDs themselves.  Regular calibration observations monitor the optical depth, composition, and spatial distribution as it has evolved over the lifetime of Chandra.\cite{ContamMeasure2,ContamMeasure}  Figure~\ref{fig:contam} demonstrates the changing effective area due to contaminant deposition.

Substantial effort has been put towards identifying the exact composition of the contaminant and its volatility.  That, along with detailed thermal models of the ACIS instrument, have been used to predict whether a bakeout of the instrument would be effective.\cite{ContamModel}  Unfortunately, the unknown contaminant volatility does not allow us to predict the results of a bakeout, how much contaminant would be removed, if any, or how fast it might reaccumulate afterwards.  A bakeout also carries some risk of damaging the Optical Blocking Filter,  so an on-orbit bakeout has not been attempted. The effectiveness and risks of a bakeout of the focal plane to remove contamination are uncertain, and the consensus is that the risks are not justified by the potential benefits.

\subsection{Particle Background} \label{sect:bkg}

Chandra's highly elliptical orbit samples a wide range of near-Earth particle environments, from the radiation belts at closest approach through the magnetosphere and magnetopause and past the bow shock into the solar wind.   Science observations take place outside the Earth's radiation belts, where the quiescent background is dominated by galactic cosmic rays (GCR), primarily protons $>$~10~MeV, plus any secondary particles produced through interaction with the spacecraft and instrument.  

\begin{figure}[t!]
\begin{center}
\includegraphics[width=6.5in]{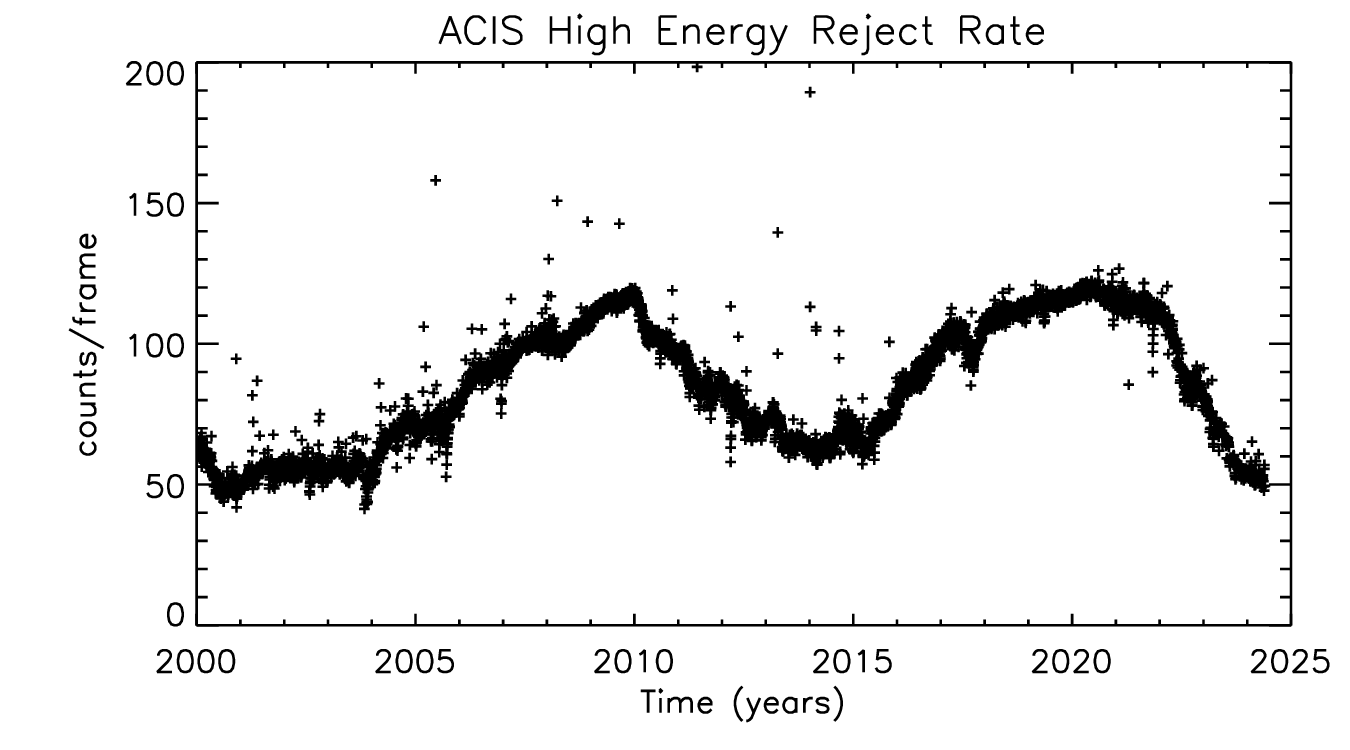}
\includegraphics[width=3in]{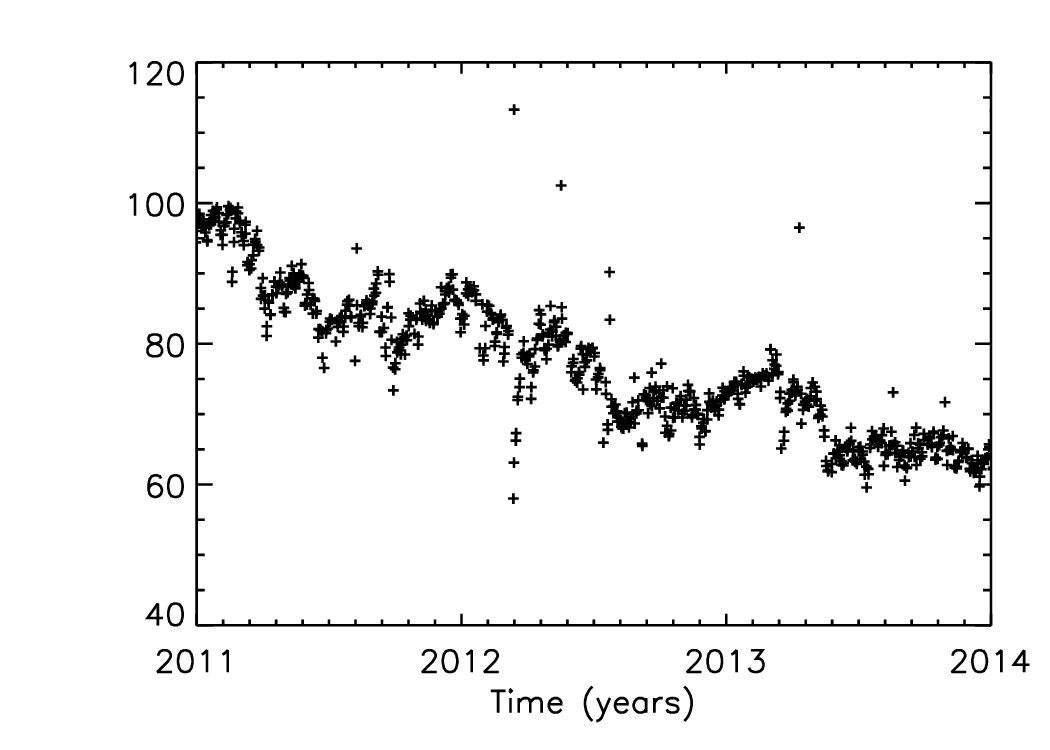}
\includegraphics[width=3in]{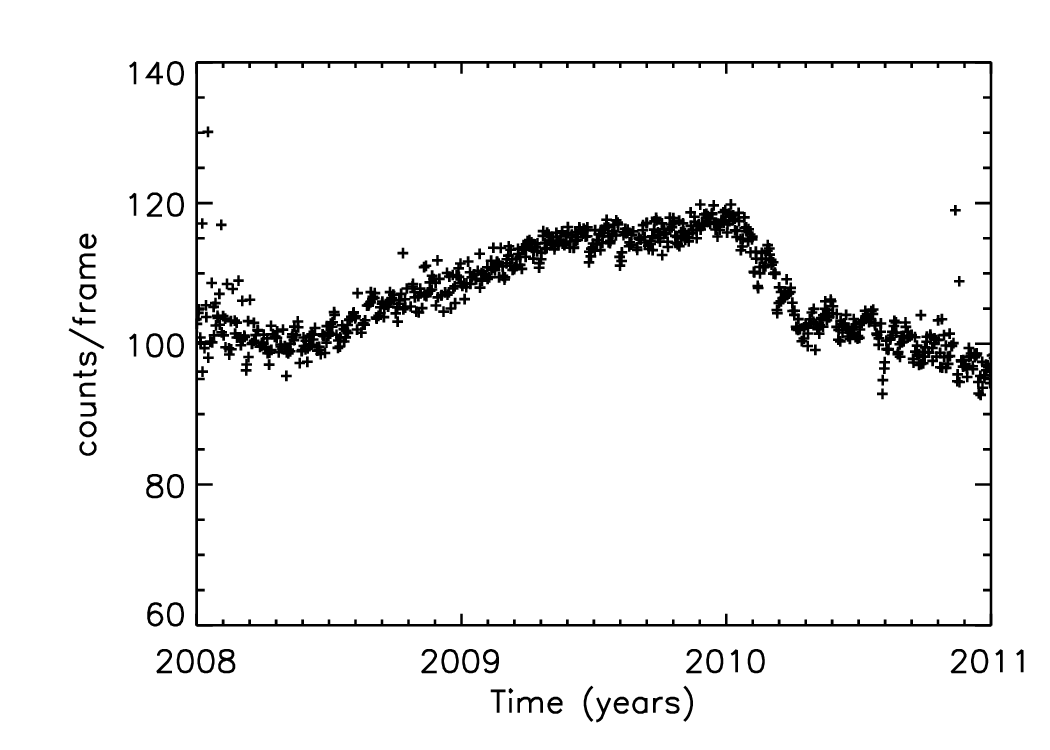}
\caption{Time dependence of the particle background as measured by the rate of high energy events on ACIS-S3. Each data point is a single ECS observation. (top) The full mission history of high energy rates on S3 includes three solar maxima and two minima. The particle background is anti-correlated with the solar cycle and varies by a factor of $\sim$2 over 11~years. The cycles are distinct; rates from one cycle could not be used to accurately predict the rates in the next cycle. (bottom left) A segment of the ACIS-S3 background during solar maximum.  The sawtooth pattern is typical of solar activity and individual solar storms. The background variation is $\sim$10\% over weeks to months. (bottom right) A segment of the ACIS-S3 background during solar minimum, when there was no significant solar activity.  Background variations are smaller, of order 5\% over a matter of days.}
\label{fig:acisbkg}
\end{center}
\end{figure}

GCR protons are the primary source of quiescent background, modulated by the solar cycle and solar activity.  The twenty-five year baseline provided by Chandra data covers more than two complete solar cycles, allowing a study of conditions at solar maximum and minimum and for variances between solar cycles.  Figure~\ref{fig:acisbkg} shows the time dependence of the particle background on ACIS using the high energy reject rate as a proxy.  Variability is present on multiple timescales.  At the largest scales, the two solar cycles can be compared.  Each one is distinct; rates from one cycle could not be used to accurately predict those in the next. During solar maximum, there is an additional $\sim 10\%$ variation on timescales of weeks to months from individual solar storms, but even at solar minimum, there remains variation at the 5\% level.

It is expected that the unfocused particle background experienced by future missions in L1 or L2 should be very similar to that seen by Chandra.  Ref.~\citenum{ACISAMS} compares the variability of ACIS to that measured by the precision particle experiment, Alpha Magnetic Spectrometer on the ISS, and to an equivalent reject rate on XMM-Newton EPIC-pn, to improve our knowledge of the background for future missions.

\subsection{Flight Software Patches}
\label{sect:txings}

The ACIS flight software is flexible and has been patched eleven times since launch.  The patches have included bug fixes, improvements to on-board processing, and enhancements to the flight software capabilities. The most recent patchload was uplinked in September 2023 with four patches: a) to correct a bug that could lead to an infinite loop during the bias map computation in alternating exposure mode, b) to accurately report the state of the software in all cases after a safing action, c) to report more diagnostic information in the case of an unexpected condition, and d) to force a restart and reloading of the software if a front-end processor is powered off unexpectedly.  Updates such as these make ACIS more operationally robust and help minimize any downtime for some anomalous conditions.  

An example of an ACIS flight software patch that has had great impact on the instrument lifetime is TXings, the ACIS radiation monitor patch. The Chandra team has implemented procedures to protect ACIS during high-radiation events including autonomous protection triggered by an on-board radiation monitor. When an elevated particle environment is reported to the On-Board Computer (OBC), science observations are stopped, the ACIS CCDs are powered down, and ACIS is translated to the stowed position.
Due to elevated temperatures, the on-board radiation monitor, EPHIN, became increasingly unreliable and in 2014 was removed from the autonomous radiation protection monitor.  The anti-coincidence shield of the HRC was added to the on-board radiation monitor in 2005, but due to operational changes after HRC anomalies in 2020 and 2022, is no longer providing that function. Concern about the effectiveness of the on-board particle monitor motivated exploring other on-board measures of the radiation environment such as ACIS itself.\cite{GrantTXing1}

The ACIS team has developed an algorithm which uses data from the CCDs themselves to detect periods of high radiation and a flight software patch to apply this algorithm is currently active on-board the instrument.\cite{GrantTXing2,FordTXing} This patch was installed on ACIS in November 2011; it examines threshold crossing rates in the ACIS CCDs and signals the OBC if the rates are above pre-set triggers levels and monotonically increasing. The OBC can then take action to protect the ACIS CCDs by moving ACIS out of the focal plane and turning off the radiation-sensitive video boards. The patch has operated as expected with no impact on regular science operations and has correctly detected an enhanced radiation environment on multiple occasions and triggered autonomous radiation protection of the spacecraft. To better match the changing quiescent particle background which is anti-correlated with the solar cycle, the parameters of the patch are re-evaluated monthly and updated as necessary.

The loss of HRC data to on-board radiation monitoring in recent years leaves the ACIS algorithm as the sole input to the autonomous radiation safing system on Chandra. ACIS and the TXings patch have limitations compared to a dedicated particle instrument, like EPHIN, the original Chandra particle detector, or the HRC anti-coincidence shield. The algorithm runs parasitically while ACIS is taking event data, not during biases or slews, and includes contributions from both astrophysical X-rays as well as particle radiation. Given these known limitations, it was decided after the second HRC anomaly to impose additional radiation conservatism. An update to the TXings patch was uplinked in 2022, which makes it more sensitive to radiation and does not solely rely on detecting a monotonically increasing signal. 

\section{Twenty-Five Years of Lessons Learned}
\label{sect:lessons}
After twenty-five years of successful on-orbit operation, there are many lessons learned that could be useful for future missions. An early example is the radiation damage suffered by the FI CCDs after radiation belt transits (Section~\ref{sect:cti}).  Once the cause was understood, the ACIS experience was communicated to the XMM-Newton team, which launched CCD instruments into a similar orbit five months later.  So warned, XMM operations included closing the filter wheel during radiation belt transits, preventing similar damage.\cite{XMMEPICMOS}  The upcoming SMILE SXI has similar plans.\cite{SMILESXI}

The contamination accumulation on ACIS (Section~\ref{sect:contam}) and on the Suzaku XIS\cite{SuzakuXIS} has motivated the design of warm contamination blocking filters planned for a number of upcoming missions including the recently launched XRISM Xtend.\cite{HitomiCBF} In both cases it is believed that the contamination was condensing on the filters rather than the CCDs.  The CCDs are generally the coldest surfaces, but are separate from the main body of the spacecraft, where more sources of contamination exist. Actively warming the filter should help keep it clean. 

The ACIS External Calibration Source (Section~\ref{sect:ecs}) has been key to the scientific success of ACIS. Fully illuminating all ten CCDs with multiple line energies has been crucial for monitoring and calibrating the spatial, spectral, and temporal aspects of radiation damage and contamination deposition.

The versatility of ACIS in reacting to anomalies and changing requirements is in part due to the ability to patch its flight software, and this in turn is greatly facilitated by the object-oriented design of its back-end processor software. Patch testing is improved by a pair of ground simulators: one in hardware built from spare flight boards, the other in software using virtual QEMU processes.

Finally, the success of ACIS operation over twenty five years is due to a large team from MIT, SAO, Penn State, Northrop Grumman, Lockheed Martin, and Marshall Space Flight Center.  In particular, the MIT instrument team and SAO science operations team work as a blended collaboration which started well before launch at the sub-assembly calibration stage. These well established relationships were crucial for the effective team response to instrument anomalies.

\section{Summary}
\label{sect:sum}

The ACIS instrument on the Chandra X-ray Observatory continues to perform well after twenty five years on-orbit with no limitations on its future operation for many years. The hardware and software are operating as designed. Increasing radiation damage, molecular contamination accumulation, and increasing temperature variation are calibration challenges but are manageable with mitigations in place. Given the lack of any replacement for Chandra's angular resolution capability, we look forward to many more years of ACIS observations.

\acknowledgments    

We are exceedingly grateful to all the scientists and engineers involved in designing, building, and supporting the continued operation of the Chandra X-ray Observatory. In particular, we would like to thank the Chandra Science Operations Team (SOT) and Flight Operations Team (FOT) at SAO, MIT, and Northrop-Grumman (NG), and the Chandra Project Science team (MSFC) for many years of fruitful collaboration and constant vigilance. This work was supported by NASA contracts NAS 8-37716 and NAS 8-38252.

\bibliography{report} 

\begin{thebibliography}{10}

\bibitem{cha2}
{Weisskopf}, M.~C., {Brinkman}, B., {Canizares}, C., {Garmire}, G., {Murray}, S., and {Van Speybroeck}, L.~P., ``{An Overview of the Performance and Scientific Results from the Chandra X-Ray Observatory},'' {\em Publications of the ASP}~{\bf 114},  1--24 (Jan. 2002).

\bibitem{ACISCCD}
{Burke}, B.~E., {Gregory}, J.~A., {Bautz}, M.~W., {Prigozhin}, G.~Y., {Kissel}, S.~E., {Kosicki}, B.~B., {Loomis}, A.~H., and {Young}, D.~J., ``{Soft-X-ray CCD imagers for AXAF.},'' {\em IEEE Transactions on Electron Devices}~{\bf 44},  1633--1642 (Oct. 1997).

\bibitem{GarmireACIS}
{Garmire}, G.~P., {Bautz}, M.~W., {Ford}, P.~G., {Nousek}, J.~A., and {Ricker}, George~R., J., ``{Advanced CCD imaging spectrometer (ACIS) instrument on the Chandra X-ray Observatory},'' in [{\em X-Ray and Gamma-Ray Telescopes and Instruments for Astronomy.}{\nolinebreak\hspace{0.1em}]},  {Truemper}, J.~E. and {Tananbaum}, H.~D., eds., {\em Society of Photo-Optical Instrumentation Engineers (SPIE) Conference Series} {\bf 4851},  28--44 (Mar. 2003).

\bibitem{Fe55decay}
{Audi}, G., {Bersillon}, O., {Blachot}, J., and {Wapstra}, A.~H., ``{The NUBASE evaluation of nuclear and decay properties},'' {\em Nuclear Physics A}~{\bf 729},  3--128 (Dec. 2003).

\bibitem{PrigozhinCTI}
{Prigozhin}, G.~Y., {Kissel}, S.~E., {Bautz}, M.~W., {Grant}, C., {LaMarr}, B., {Foster}, R.~F., and {Ricker}, G.~R., ``{Characterization of the radiation damage in the Chandra x-ray CCDs},'' in [{\em X-Ray and Gamma-Ray Instrumentation for Astronomy XI}{\nolinebreak\hspace{0.1em}]},  {Flanagan}, K.~A. and {Siegmund}, O.~H., eds., {\em Society of Photo-Optical Instrumentation Engineers (SPIE) Conference Series} {\bf 4140},  123--134 (Dec. 2000).

\bibitem{TownsleyCTIcorr}
{Townsley}, L.~K., {Broos}, P.~S., {Nousek}, J.~A., and {Garmire}, G.~P., ``{Modeling charge transfer inefficiency in the Chandra Advanced CCD Imaging Spectrometer},'' {\em Nuclear Instruments and Methods in Physics Research A}~{\bf 486},  751--784 (July 2002).

\bibitem{GrantCTIcorr}
{Grant}, C.~E., {Bautz}, M.~W., {Kissel}, S.~E., and {LaMarr}, B., ``{A charge transfer inefficiency correction model for the Chandra advanced CCD imaging spectrometer},'' in [{\em High-Energy Detectors in Astronomy}{\nolinebreak\hspace{0.1em}]},  {Holland}, A.~D., ed., {\em Society of Photo-Optical Instrumentation Engineers (SPIE) Conference Series} {\bf 5501},  177--188 (Sept. 2004).

\bibitem{OdellCTI}
{O'Dell}, S.~L., {Aldcroft}, T.~L., {Blackwell}, W.~C., {Bucher}, S.~L., {Chappell}, J.~H., {DePasquale}, J.~M., {Grant}, C.~E., {Juda}, M., {Martin}, E.~R., {Minow}, J.~I., {Murray}, S.~S., {Plucinsky}, P.~P., {Schwartz}, D.~A., {Shropshire}, D.~P., {Spitzbart}, B.~J., {Viens}, P.~R., and {Wolk}, S.~J., ``{Managing radiation degradation of CCDs on the Chandra X-ray Observatory III},'' in [{\em UV, X-Ray, and Gamma-Ray Space Instrumentation for Astronomy XV}{\nolinebreak\hspace{0.1em}]},  {Siegmund}, O.~H., ed., {\em Society of Photo-Optical Instrumentation Engineers (SPIE) Conference Series} {\bf 6686},  668603 (Sept. 2007).

\bibitem{GrantCTI}
{Grant}, C.~E., {Bautz}, M.~W., {Kissel}, S.~M., {LaMarr}, B., and {Prigozhin}, G.~Y., ``{Long-term trends in radiation damage of Chandra x-ray CCDs},'' in [{\em UV, X-Ray, and Gamma-Ray Space Instrumentation for Astronomy XIV}{\nolinebreak\hspace{0.1em}]},  {Siegmund}, O. H.~W., ed., {\em Society of Photo-Optical Instrumentation Engineers (SPIE) Conference Series} {\bf 5898},  201--211 (Aug. 2005).

\bibitem{CTItemp}
{Grant}, C.~E., {Bautz}, M.~W., {Kissel}, S.~E., {LaMarr}, B., and {Prigozhin}, G.~Y., ``{Temperature dependence of charge transfer inefficiency in Chandra X-ray CCDs},'' in [{\em High Energy, Optical, and Infrared Detectors for Astronomy II}{\nolinebreak\hspace{0.1em}]},  {Dorn}, D.~A. and {Holland}, A.~D., eds., {\em Society of Photo-Optical Instrumentation Engineers (SPIE) Conference Series} {\bf 6276},  62761O (June 2006).

\bibitem{ContamMeasure2}
{Marshall}, H.~L., {Tennant}, A., {Grant}, C.~E., {Hitchcock}, A.~P., {O'Dell}, S.~L., and {Plucinsky}, P.~P., ``{Composition of the Chandra ACIS contaminant},'' in [{\em X-Ray and Gamma-Ray Instrumentation for Astronomy XIII}{\nolinebreak\hspace{0.1em}]},  {Flanagan}, K.~A. and {Siegmund}, O. H.~W., eds., {\em Society of Photo-Optical Instrumentation Engineers (SPIE) Conference Series} {\bf 5165},  497--508 (Feb. 2004).

\bibitem{ContamMeasure}
{Plucinsky}, P.~P., {Bogdan}, A., and {Marshall}, H.~L., ``{The evolution of the ACIS contamination layer on the Chandra X-ray Observatory through 2022},'' in [{\em Space Telescopes and Instrumentation 2022: Ultraviolet to Gamma Ray}{\nolinebreak\hspace{0.1em}]},  {den Herder}, J.-W.~A., {Nikzad}, S., and {Nakazawa}, K., eds., {\em Society of Photo-Optical Instrumentation Engineers (SPIE) Conference Series} {\bf 12181},  121816X (Aug. 2022).

\bibitem{ContamModel}
{O'Dell}, S.~L., {Swartz}, D.~A., {Tice}, N.~W., {Plucinsky}, P.~P., {Marshall}, H.~L., {Bogdan}, A., {Grant}, C.~E., {Tennant}, A.~F., and {Dahmer}, M., ``{Modeling contamination migration on the Chandra X-ray Observatory IV},'' in [{\em Society of Photo-Optical Instrumentation Engineers (SPIE) Conference Series}{\nolinebreak\hspace{0.1em}]},  {Siegmund}, O.~H., ed., {\em Society of Photo-Optical Instrumentation Engineers (SPIE) Conference Series} {\bf 10397},  103970C (Aug. 2017).

\bibitem{ACISAMS}
{Sarkar}, A., {Grant}, C.~E., {Miller}, E.~D., {Bautz}, M., {Schneider}, B., {Foster}, R.~F., {Schellenberger}, G., {Allen}, S., {Kraft}, R.~P., {Wilkins}, D., {Falcone}, A., and {Ptak}, A., ``{Advancing Precision Particle Background Estimation for Future X-ray Missions: Correlated Variability between AMS and Chandra/XMM-Newton},'' {\em Astrophysical Journal}~{\bf Submitted},  arXiv:2405.06602 (2024).

\bibitem{GrantTXing1}
{Grant}, C.~E., {Lamarr}, B., {Bautz}, M.~W., and {O'Dell}, S.~L., ``{Using ACIS on the Chandra X-ray Observatory as a particle radiation monitor},'' in [{\em Space Telescopes and Instrumentation 2010: Ultraviolet to Gamma Ray}{\nolinebreak\hspace{0.1em}]},  {Arnaud}, M., {Murray}, S.~S., and {Takahashi}, T., eds., {\em Society of Photo-Optical Instrumentation Engineers (SPIE) Conference Series} {\bf 7732},  77322I (July 2010).

\bibitem{GrantTXing2}
{Grant}, C.~E., {Ford}, P.~G., {Bautz}, M.~W., and {O'Dell}, S.~L., ``{Using ACIS on the Chandra X-ray Observatory as a particle radiation monitor II},'' in [{\em Space Telescopes and Instrumentation 2012: Ultraviolet to Gamma Ray}{\nolinebreak\hspace{0.1em}]},  {Takahashi}, T., {Murray}, S.~S., and {den Herder}, J.-W.~A., eds., {\em Society of Photo-Optical Instrumentation Engineers (SPIE) Conference Series} {\bf 8443},  844311 (Sept. 2012).

\bibitem{FordTXing}
{Ford}, P.~G. and {Grant}, C.~E., ``{Using the Chandra ACIS x-ray imager as a background particle flux detector},'' in [{\em Space Telescopes and Instrumentation 2012: Ultraviolet to Gamma Ray}{\nolinebreak\hspace{0.1em}]},  {Takahashi}, T., {Murray}, S.~S., and {den Herder}, J.-W.~A., eds., {\em Society of Photo-Optical Instrumentation Engineers (SPIE) Conference Series} {\bf 8443},  844347 (Sept. 2012).

\bibitem{XMMEPICMOS}
{Sembay}, S., {Abbey}, A., {Altieri}, B., {Ambrosi}, R., {Baskill}, D., {Ferrando}, P., {Mukerjee}, K., {Read}, A.~M., and {Turner}, M. J.~L., ``{In-orbit performance of the EPIC-MOS detectors on XMM-Newton},'' in [{\em UV and Gamma-Ray Space Telescope Systems}{\nolinebreak\hspace{0.1em}]},  {Hasinger}, G. and {Turner}, M. J.~L., eds., {\em Society of Photo-Optical Instrumentation Engineers (SPIE) Conference Series} {\bf 5488},  264--271 (Oct. 2004).

\bibitem{SMILESXI}
{Sembay}, S., {Alme}, A.~L., {Agnolon}, D., {Arnold}, T., {Beardmore}, A., {Bel{\'e}n Balado Margeli}, A., {Bicknell}, C., {Bouldin}, C., {Branduardi-Raymont}, G., {Crawford}, T., {Breuer}, J.~P., {Buggey}, T., {Butcher}, G., {Canchal}, R., {Carter}, J.~A., {Cheney}, A., {Collado-Vega}, Y., {Connor}, H., {Crawford}, T., {Eaton}, N., {Feldman}, C., {Forsyth}, C., {Frantzen}, T., {Galg{\'o}czi}, G., {Garcia}, J., {Genov}, G.~Y., {Gordillo}, C., {Gr{\"o}belbauer}, H.~P., {Guedel}, M., {Guo}, Y., {Hailey}, M., {Hall}, D., {Hampson}, R., {Hasiba}, J., {Hetherington}, O., {Holland}, A., {Hsieh}, S.~Y., {Hubbard}, M.~W.~J., {Jeszenszky}, H., {Jones}, M., {Kennedy}, T., {Koch-Mehrin}, K., {K{\"o}gl}, S., {Krucker}, S., {Kuntz}, K.~D., {Laky}, G., {Lylund}, O., {Martindale}, A., {Miguel Mas Hesse}, J., {Nakamura}, R., {Oksavik}, K., {{\O}stgaard}, N., {Ottacher}, H., {Ottensamer}, R., {Pagani}, C., {Parsons}, S., {Patel}, P., {Pearson}, J., {Peikert}, G., {Porter}, F.~S., {Pouliantis}, T., {Qureshi}, B.~H., {Raab},
  W., {Randall}, G., {Read}, A.~M., {Roque}, N.~M.~M., {Rostad}, M.~E., {Runciman}, C., {Sachdev}, S., {Samsonov}, A., {Soman}, M., {Sibeck}, D., {Smit}, S., {S{\o}ndergaard}, J., {Speight}, R., {Stavland}, S., {Steller}, M., {Sun}, T., {Thornhill}, J., {Thomas}, W., {Ullaland}, K., {Walsh}, B., {Walton}, D., {Wang}, C., and {Yang}, S., ``{The Soft X-ray Imager (SXI) on the SMILE Mission},'' {\em Earth and Planetary Physics}~{\bf 8},  5--14 (Jan. 2024).

\bibitem{SuzakuXIS}
{Koyama}, K., {Tsunemi}, H., {Dotani}, T., {Bautz}, M.~W., {Hayashida}, K., {Tsuru}, T.~G., {Matsumoto}, H., {Ogawara}, Y., {Ricker}, G.~R., {Doty}, J., {Kissel}, S.~E., {Foster}, R., {Nakajima}, H., {Yamaguchi}, H., {Mori}, H., {Sakano}, M., {Hamaguchi}, K., {Nishiuchi}, M., {Miyata}, E., {Torii}, K., {Namiki}, M., {Katsuda}, S., {Matsuura}, D., {Miyauchi}, T., {Anabuki}, N., {Tawa}, N., {Ozaki}, M., {Murakami}, H., {Maeda}, Y., {Ichikawa}, Y., {Prigozhin}, G.~Y., {Boughan}, E.~A., {Lamarr}, B., {Miller}, E.~D., {Burke}, B.~E., {Gregory}, J.~A., {Pillsbury}, A., {Bamba}, A., {Hiraga}, J.~S., {Senda}, A., {Katayama}, H., {Kitamoto}, S., {Tsujimoto}, M., {Kohmura}, T., {Tsuboi}, Y., and {Awaki}, H., ``{X-Ray Imaging Spectrometer (XIS) on Board Suzaku},'' {\em Publications of the ASJ}~{\bf 59},  23--33 (Jan. 2007).

\bibitem{HitomiCBF}
{Kohmura}, T., {Kaneko}, K., {Tsunemi}, H., {Hayashida}, K., {Nagino}, R., {Inoue}, S., {Uchida}, D., {Katada}, S., {Dotani}, T., {Ozaki}, M., {Tomida}, H., {Kimura}, M., {Tsuru}, T.~G., {Ikeda}, S., {Yabe}, K., {Miyakawa}, K., {Andoh}, M., {Kuwano}, S., {Sato}, Y., {Tamasawa}, K., {Tanno}, S., and {Yoshino}, Y., ``{Soft x-ray transmission of contamination blocking filter for SXI onboard ASTRO-H},'' in [{\em Space Telescopes and Instrumentation 2014: Ultraviolet to Gamma Ray}{\nolinebreak\hspace{0.1em}]},  {Takahashi}, T., {den Herder}, J.-W.~A., and {Bautz}, M., eds., {\em Society of Photo-Optical Instrumentation Engineers (SPIE) Conference Series} {\bf 9144},  91445D (July 2014).

\end{thebibliography}
\bibliographystyle{spiebib} 

\end{document}